\documentstyle[preprint,aps,pre]{revtex}

\begin{document}

\title{Statistical properties of genealogical trees}

\author{Bernard Derrida$^1$, Susanna C. Manrubia$^2$, and
Dami\'an H. Zanette$^3$}

\address{
$^1$ Laboratoire de Physique Statistique de l'\'Ecole Normale
Sup\'erieure\\ 24 rue Lhomond, F-75231 Paris 05 Cedex, France\\
$^2$ Fritz-Haber-Institut der Max-Planck-Gesellschaft\\
Faradayweg 4-6, 14195 Berlin, Germany\\
$^3$ Consejo  Nacional de  Investigaciones  Cient\'{\i}ficas  y
T\'ecnicas\\ Centro At\'omico Bariloche  e  Instituto  Balseiro\\
8400  S.C.  de Bariloche, R\'{\i}o Negro, Argentina}

\date{\today}

\maketitle

\begin{abstract}
We  analyse  the  statistical  properties  of  genealogical trees in 
a neutral model of a
closed  population  with   sexual  reproduction  and   non-overlapping
generations.  By reconstructing the genealogy 
of an individual from the population
evolution, we  measure the distribution  of 
ancestors appearing  more  than  once  in  a  given  tree. 
After a transient time, the  probability  of
repetition follows, up to a rescaling,  a stationary distribution which we calculate
both numerically and analytically. This distribution exhibits a
universal  shape with  a  non-trivial power law which can be understood
by   an exact, though simple, 
renormalization calculation. Some real data on  human genealogy
illustrate the  problem, which  is relevant  to the  study of the real
degree of diversity in closed interbreeding communities.

\end{abstract}


\newpage

\baselineskip=5mm

Modern man appeared on Earth some $10^5$ years ago \cite{Brau,Cavalli}. 
At that time, few  social  groups totalling several  thousands   of
individuals were occupying small settlements, most probably in Africa
\cite{Reich}.
Nowadays, we  are faced  with about  $5 \times  10^9$ human  beings on
Earth, whose lineages could be in principle traced back to that  time.
Each human being  has two parents,  four grandparents, and  in general
$2^{n_g}$ ancestors in the $n_g-$th upper generation. Going  backwards
in time  until the  first group  of anatomically modern 
{\it  Homo sapiens} --some
4000  generations  ago--  we  should  find  $2^{4000}\sim   10^{1200}$
ancestors  in  each  genealogical  tree.  However,  the  total   human
population  at  those  early  times  was probably
of a few thousands only! The
answer to this apparent paradox is simple: a given individual  appears
more than once in a genealogical tree  \cite{Ohno},
 even in very distant  branches,
indicating that many  of the ancestors  were in fact  close relatives.
A repeated  individual generates a whole
repeated branch in the tree, and the further we  move into
the  past,  the  more  frequent  the  repetitions will be. This is the
result of mating inside a finite population, the size of which sets an
upper bound to the maximal number of ancestors for a given individual.

These repetitions are particularly apparent  when we are faced with  a
small closed  interbreeding population.   Royal genealogy  provides us
with a  nice example,  since nobles  usually married  within their own
castes.  As  an  illustration to the  problem,  we  have analysed the
repetitions in the  genealogical tree of  the English king  Edward III
(1312-1377) \cite{Edw}. It contains almost $10^3$ individuals, some of
which appear more  than once (and  up to six  times) in his  tree. 
We show in Fig. 1 the function $F(r)$, defined as the quotient between 
the number $M(r)$ of ancestors which appear $r$ times in the tree and 
the total number of different ancestors $N_t$, $F(r)=M(r)/N_t$.

We study here the statistics of repetitions in genealogical trees
as a function of the population size and the generation in the past 
that we are looking at. The question that we are addressing can
be put in the more general context of genetic diversity \cite{Harp,Wiuf}.
In fact, an important factor in the variability of natural populations
is the diversity displayed, in the genealogical history
of every individual, by his ancestors themselves, and by their weights 
in the present genome.
 Here we calculate these weights in a simple neutral model, with no selection, no change
in the population size and
  no geographical isolation.   Possible effects of these on
genealogies and genetic diversities are discussed in \cite{Ohno,Boeh,Tish}.

We  have  started  by  performing  numerical  simulations  of a 
simple neutral model of a closed
population  evolving  under  sexual  reproduction with non-overlapping
generations. In our model the population size 
is fixed to be $N$ for all generations. The population is equally 
divided into two  groups, representing  males  and  females.  At  every
generation,  we form heterosexual pairs at random and assign them a 
certain number  of descendants according to a Poisson distribution. 
This is done by choosing for each male or each female a pair of parents
at random in the previous generation \cite{Serva}. After a number  $G$ 
of generations, the  tree of each of  the
individuals  in  the   youngest  generation  is reconstructed. 

We have first calculated  the   
distribution $F(r)$ of  repetitions in this model
  for  a  population    $N=2^{11}$  and
$N=2^{12}$ individuals. This might be  a rough estimate of the  number
of noble people at the time  of Edward III. After $G=10$ generations,  we
compute the probability of repetitions in the whole tree (notice  that
in the real world generations often overlap and thus the same person might
be found  in different  generations;  this possibility is absent in our 
model).  The result of our simulations is  compared with the
real data displayed  in Figure 1. We observe  an acceptable agreement,
although we should say that the distribution $F(r)$ depends rather  strongly
on $G$ and $N$  and that the agreement is
often worse for other reasonable choices of these two parameters.

We have also  measured the
probability  of  repetitions  $H(r,n_g)$  at every past generation 
$n_g=1, \dots, G$, that is,  the  probability that any  individual
at generation $n_g$ in the past appears $r$  times in the tree of an individual
at generation $0$ ($n_g=1$ corresponds to the parents, $n_g=2$ to the grandparents 
and so on; note  that $ H(0,n_g) $ is  simply the probability that an individual  
is not present in a tree after $n_g$ generations). In the  first few  generations 
(parents, grandparents...), if the population size $N$ is large,
the probability of finding an individual more than once in a tree is 
very small. As a consequence $H(r,n_g)$ decreases  with $r$ when $n_g$ is small.
Going further in the past, at  some  point  two  ``brothers"  will 
appear in the tree of an individual, and from then on these two 
branches will coincide.  From then on, more and more repetitions will occur.

The distribution $H(r,n_g)$ is shown in Figure 2. It changes its shape  during a transient period
of the  order  of  $\log  N$  generations.  
(Note that an important difference between $F(r)$ shown in figure 1 and $H(r,n_g)$ is that for $F$
we counted only those individuals present in one particular genealogical tree
whereas for $H$ we count the whole population at generation $n_g$  in the past.)
Clearly, we have $\sum_{r \geq 0} H(r,n_g)= 1$
and $\sum_r r H(r,n_g) = 2^{n_g}$. In figure 2 we see
  for $N=2^{15}$ the function $H(r,n_g)$ for different generations 
before and after reaching the stationary shape.
For $n_g$ small, $H(r,n_g)$ decreases with $r$, meaning that repetitions are
very unprobable. As $n_g$ increases, the number of repetitions increases
and $H(r,n_g)$ exhibits a maximum and a shape which becomes  stationary.

 If we rescale the distribution $H(r,n_g)$ by
plotting as in Figure 3 the distribution 
\begin{equation}
P(w) \equiv  2^{n_g} H(r,n_g)/N  
\label{Pw}
\end{equation}
versus  the weight
\begin{equation}
w \equiv  r N/2^{n_g}
\label{w}
\end{equation}
all the distributions of Figure 2 (after a transient period)   collapse on a 
single  stationary function.
Figure 3 represents  the function $P(w)$  for several values  of $n_g$
after the transient period
 obtained for  a population of  $N=2^{20}$ individuals. 
We observe that the 
left tail of $P(w)$ is a power law, $P(w) \sim w^{\beta}$, and a least 
squares fit to  our  numerical  results  in  the  domain $w \in 
(10^{-4},10^{-1})$ returns $\beta \simeq 0.302$. In addition to the exponent 
$\beta$, one can accurately measure the moments of $\langle w^n \rangle 
= \int w^n P(w) dw $ of $P(w) $ as well as the fraction $S(n_g)$ of the  total
population in the oldest generation    which is absent from a given  genealogical tree.
 Figure 4 contains our numerical estimates for $S(n_g)$.
Figure 4 shows also
 the first moments of the distribution $P(w)$. As can 
be seen, even when the number of potential ancestors in the tree is much 
larger than the number of individuals in the population, not all of
those give contributions to the present. 
In fact, the proportion of individuals without descendents 
reaches a fixed value, $S(n_g \to \infty) \simeq 0.2031$. 

The distribution $P(w)$ can be understood analytically by the following 
argument: if we consider the genealogical tree of an individual, say 
individual $i=1$ at the $0$th generation, the weights $w$  of 
his ancestors can be traced back according to the following algorithm. 
 From (\ref{Pw},\ref{w}) we have $w_i(0)=N$ for 
$i=1$ and $w_i(0)=0$ for $i \neq 1$. Then the weights of the ancestors 
at  generation $n_g+1$ in the past, with $ 0 \leq n_g \leq G-1$, are 
given by
\begin{equation}
w_i(n_g+1) = \sum_{ j \makebox{ children   of } i}  
{ w_j(n_g)  \over 2}
\label{eq1}
\end{equation}

When $N$ is large, the probability $p_k $ that an individual
at generation $n_g +1 $ in the past  has $k$ children at generation $n_g$
becomes a Poisson distribution
\begin{equation}
p_k = 2^k e^{-2} / k!
\label{eq2}
\end{equation}

Now if for large $N$ we consider 
that the weights of the children of any given individual are uncorrelated,
(this can be viewed as an approximation, but in fact, by calculating pair 
correlations between the weights in our model, one can show that for 
large enough $N$
this approximation becomes exact), we obtain from  (\ref{eq1},\ref{eq2}) 
that any weight at generation $n_g +1$ is the sum of $k$ i.i.d. weights 
at generation $n_g$ with $k$ itself randomly chosen according to 
(\ref{eq2}). Then if $g_{n_g}(\lambda)$ is the generating function of the 
weights at generation $n_g$ in the past,
$$ g_{n_g}(\lambda) = \left\langle e^{\lambda w_i(n_g)} \right\rangle$$
it follows from (3) (and the fact that  for large $N$ the $w_j(n_g)$ are 
uncorrelated) that it satisfies
\begin{equation}
 g_{n_g+1}(\lambda) = \sum_{k=0}^\infty {2^k e^{-2} \over k!} \left[g_{n_g}(\lambda/2)\right]^k = e^{-2+2 g_{n_g }(\lambda /2) }
\label{eq5}
\end{equation}
 This recursion has the form of a renormalization group transformation.
 Together with the initial condition 
\begin{equation}
g_0(\lambda) =1 + ( e^{\lambda N} -1 )/N 
\label{eq4}
\end{equation}
it determines all the generating functions $g_{n_g}(\lambda)$.
When $n_g \to \infty$, the generating function converges to a limit 
$g(\lambda)$ solution of 
\begin{equation}
g(\lambda) = e^{2 g(\lambda/2) -2}.
\label{eq3}
\end{equation}

All the informations on  the shape of  the stationary solution $P(w)$ are 
contained in the solution of (\ref{eq3}). For example, one can expand 
$g(\lambda)$ solution of (\ref{eq3}) in power series and find that
$$g(\lambda) = 1 + \lambda +  \lambda^2 + {8  \over 9} \lambda^3 +
 { 46 \over 63 }\lambda^4 +{2672 \over 4725}  \lambda^5 + { 183712 \over 439425}
 \lambda^6 + ... $$
This leads to
$\langle w \rangle =1$,
$\langle w^2 \rangle =2$,
$\langle w^3 \rangle =16/3$,
$\langle w^4 \rangle =368/21$ and so on. (Note that $\langle w \rangle=1$
is not determined by (\ref{eq3}) but this is an immediate consequence of 
the initial condition (\ref{eq4}).) One can also determine the fraction 
$S$ of individuals with no descendence (that is the probability that 
$w=0$) by $S=g(-\infty)$. Clearly, $S=g(-\infty)$ is  the solution of
$$S= e^{2 S-2} $$
and this gives $S=.20318787..$.

The power law   $P(w) \sim w^{\beta}$ at small $w$ can also be easily 
understood from (\ref{eq3}). If $P(w) \simeq A w^{\beta}$ for small $w$, 
one can write that as $\lambda \to - \infty$ 
$$g(\lambda) - S \simeq A \Gamma(\beta +1) |\lambda|^{-\beta-1}$$ and   
equation (\ref{eq3}) gives (by the standard renormalization argument 
used to calculate exponents by linearization around a fixed point and 
which consists in writing the compensation of the singularities 
proportional to $|\lambda|^{-\beta-1}$ on both sides of (\ref{eq3}))
that 
$$ \beta= -{\log S \over \log2} -2 \simeq .2991138...$$
in excellent agreement with the results of the simulation.
Other properties of the stationary distribution $P(w)$ could in 
principle be extracted from (\ref{eq3}) but this would require more 
complicated mathematical developments.

In this work, we have shown that a simple neutral model of sexual 
reproduction with non-overlapping generations leads to a universal 
distribution of the weights of ancestors in genealogical trees. This 
universal distribution (more precisely its generating function) is the 
fixed point (\ref{eq3}) of a simple renormalization equation (\ref{eq5}). 
The exponent $\beta$ of the power law observed for small weights can be 
calculated exactly.

Our main result is that if we  go very far in the past, 
about 80\% of the (adult) population appears in
    the genealogical tree of every individual. If  the weights
    of these ancestors represent how often they appear in this tree,
    these weights  have a  stationary  probability distribution
    which is universal (i.e. independent  of the generation and of
    the population size).

There are a number of extensions of the present work which, in our 
opinion, are worth pursuing. First, a more complete description of $P(w)$, 
in particular the large $w$ behavior, could be extracted from (\ref{eq3}). 
If we wish to perform a better approximation to  real genealogy, 
the possibility of overlaps between generations or of changes in 
the population size should be included. 
One can try to measure the distribution of lengths of
segments in simple  models \cite{Wiuf,JM} for the evolution of chromosomes
to see whether a power law in the length distribution is present there 
too. One could also investigate how our results would be modified
by choosing instead of (\ref{eq2})  a non-Poissonian distribution of
offsprings. Lastly, it would be interesting to consider the genealogical trees 
of several individuals to see how the repetitions on different  trees are 
correlated \cite{DMZ}.

With a little more imagination, one can construct other universality 
classes, by allowing the number $p$ of parents of each individual to be 
arbitrary, instead of $p=2$ in our earthy world. For general 
$p$, the fixed point equation (\ref{eq3}) would become $g(\lambda)= 
\exp[-p+pg(\lambda/p)]$. No need to say that one might then try to 
expand the distribution $P(w)$, the fixed point $S$ or the exponent 
$\beta$ in powers of $\epsilon$ for $p=1+\epsilon$.
In fact, one can show \cite{DMZ} that the case of an exponentially 
increasing (or decreasing) population size with $p=2$ parents for each
individual is equivalent, as long as $g(\lambda)$ is concerned, to the 
case of a population of constant size with a number of parents $p$ which 
depends on the exponential growth rate of the population.

Apart from the potential application of our results to population genetics
and evolutionary biology,
the model of evolution studied here is connected to a number of
problems of current interest in physics. First, the random assignment of 
the parents of individuals at each generation is very reminiscent of a 
problem of repartition of constraints  introduced recently  \cite{Copper} 
to describe granular materials, with  a recursion similar to (\ref{eq1}).
Graphs which locally  look like trees but where large loops --responsible 
for cooperative effects-- are present have  attracted a lot of interest 
in the theory of disordered systems (spin glasses, localization)
\cite{SG1}-\cite{SG5},\cite{Loc}.

Lastly, the model studied here gives through (\ref{eq5},\ref{eq3})
a very simple and pedagogical example of a  problem with a non-trivial 
exponent, which can be solved 
exactly by  a discrete renormalization transformation. One could try to 
see whether the oscillations \cite{Sornette} which usually accompany  
such discrete renormalization  transformations are present here too.

\section*{Acknowledgements}

Interesting discussions with Ugo Bastolla are gratefully acknowledged.
SCM acknowledges support from the Alexander von Humboldt Foundation (Germany)
and from Fundaci\'on Antorchas (Argentina).

\newpage

\section*{Figure caption}

\begin{enumerate}

\item {Probability of ancestor repetitions in the genealogical tree
of the king Edward III \cite{Edw}. The continuous and dashed lines
represent the results of simulations  of $F(r)$
 in a closed population with
$2^{11}$ and $2^{12}$ individuals for our model. Averages have been
performed over the 10 first generations of $10^3$ independent trees.}

\item {Distribution $H(r,n_g)$ of $r$ repetitions after $n_g$ generations
($H(0,n_g)$ is not shown).
The distribution changes after roughly $\log N$ generations  from a decreasing function  of $r$ to a distribution
with a maximum. 
The generations shown are $n_g=9, \; 13, \; 15, \; 17, \;
19, \; 21,$ and $23$ for a population with $N=2^{15}$. We have averaged
over 100 independent runs.}

\item {Data collapse for the rescaled distribution of repetitions $P(w)$ 
after the transient period.  Averages have been performed over $10^3$ 
independent trees for a population size $N=2^{20}$.}

\item {Dependence of $S(n_g)$ on the generation $n_g$ for a population with
$N=2^{15}$. The numerical asymptotic 
value is $S(n_g \to \infty) \simeq 0.2031$. The bold dotted line is
the  predicted theoretical value $S=g(-\infty)=0.20318787 \dots$.
In the inset, we represent 
the first ten moments $\langle w^n \rangle$ for the distribution 
$P(w)$. The continuous line corresponds to numerical results, while 
solid circles stand for the theoretical predictions.}

\end{enumerate}

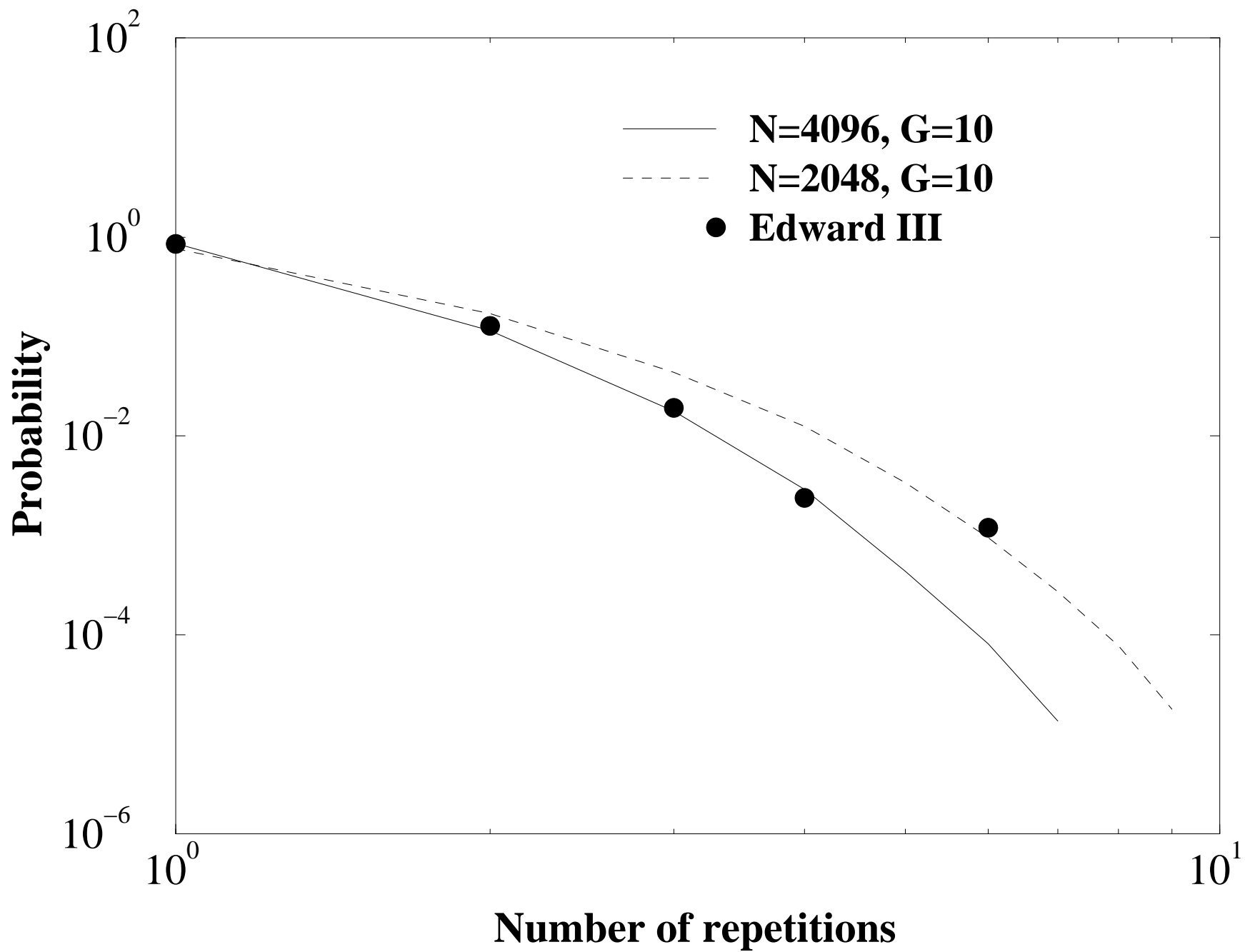
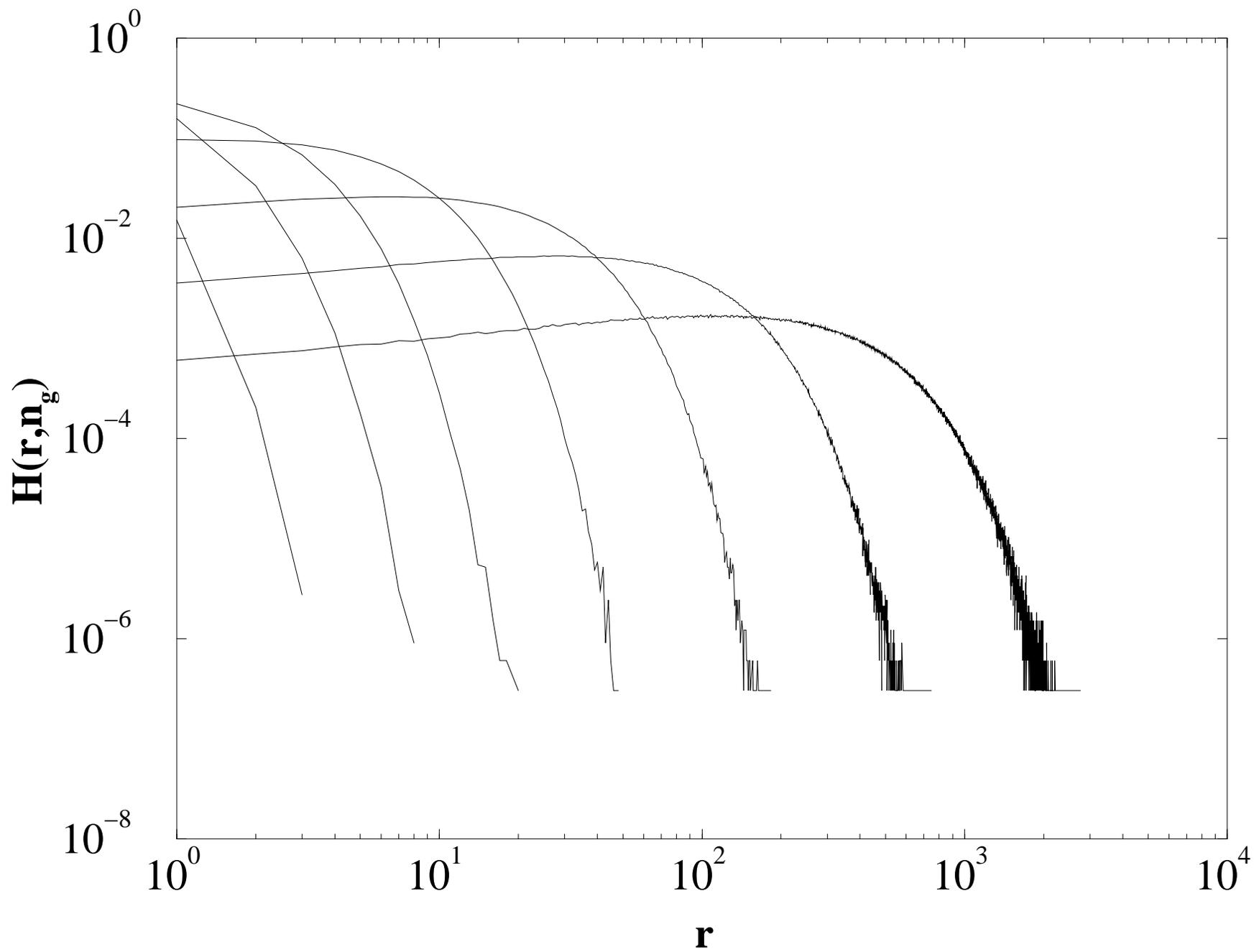
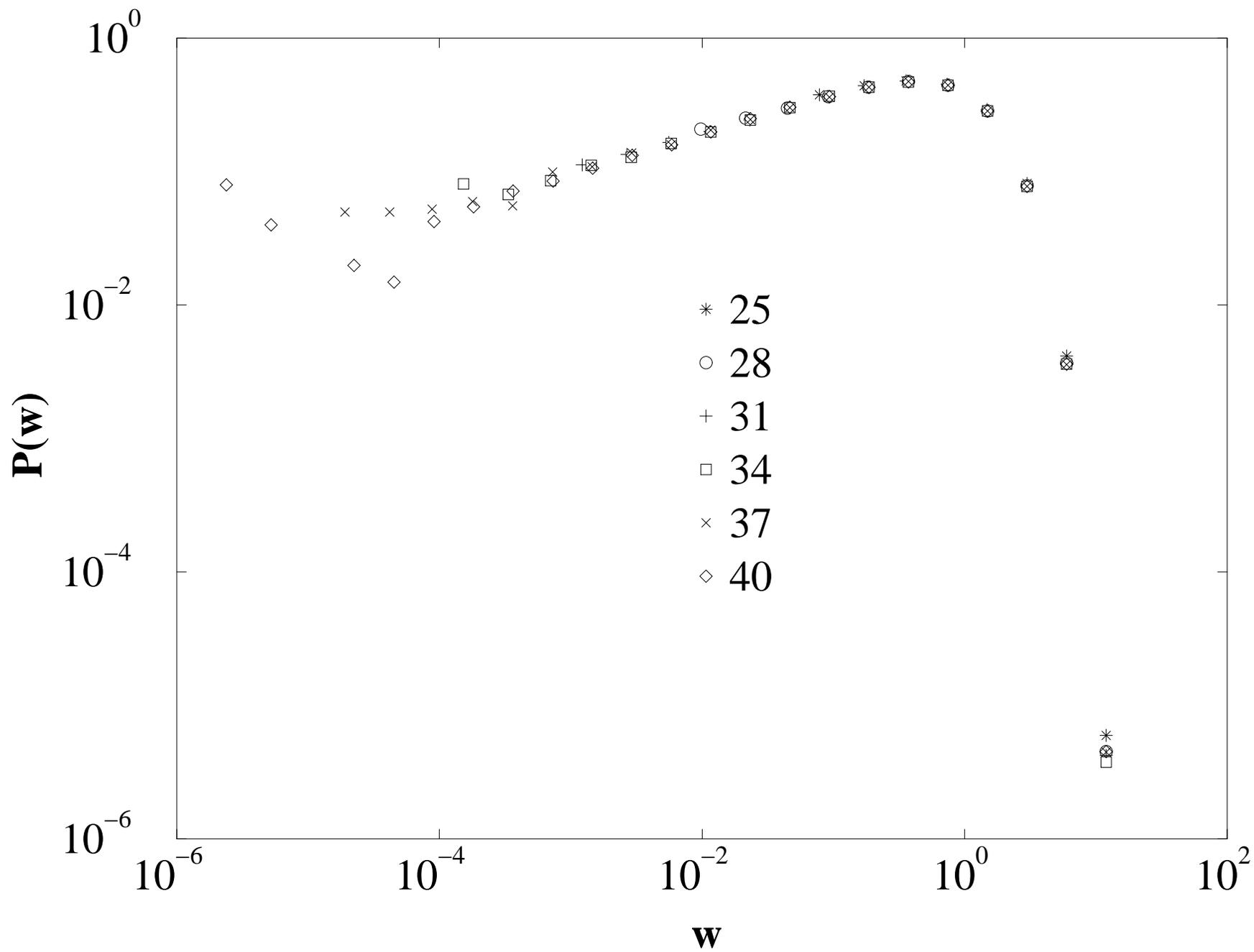
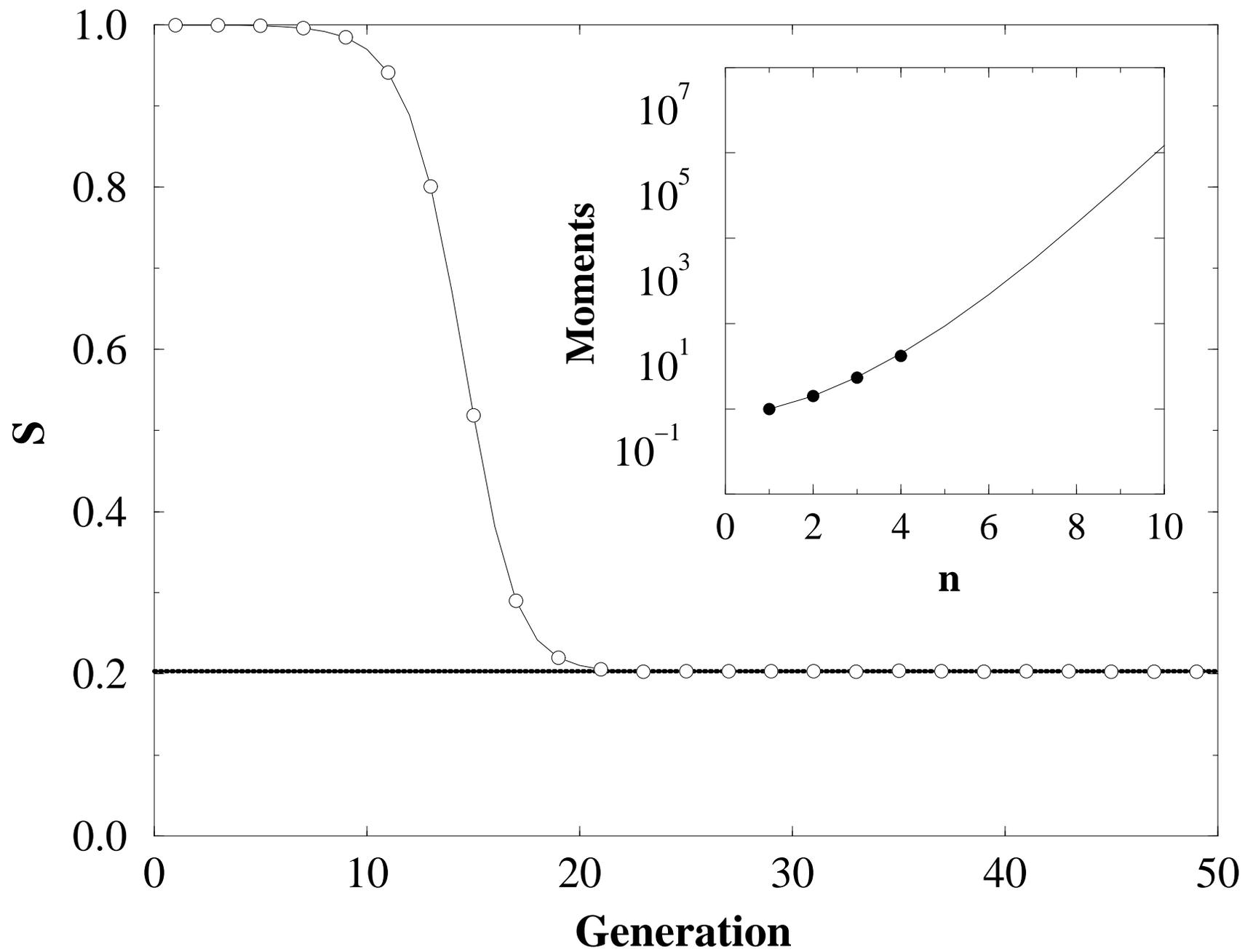
\end{document}